\documentclass[11pt,xdvi]{article}

\usepackage{graphicx}

\newcommand{\Phibar}{\overline{\delta\Phi}}
\newcommand{\eq}{Eq.}
\newcommand{\fig}{Fig.}
\newcommand{\figs}{Figs.}

\begin{document}

\title{Relating magnetic reconnection to coronal heating}


\author{
D.W.\ Longcope$^{1}$, L.A.\ Tarr$^{1,2}$
\\$^{1}$Montana State University, Bozeman, MT USA 59717\\
$^{2}$HAO/NCAR PO Box 3000, Boulder, CO USA 80307}

\maketitle

\begin{abstract}
It is clear that the solar corona is being heated and that coronal magnetic fields undergo reconnection all the time.  Here we attempt to show that these two facts are in fact related --- i.e.\ coronal reconnection generates heat.  This attempt must address the fact that topological change of field lines does not automatically generate heat.  We present one case of flux emergence where we have measured the rate of coronal magnetic reconnection and the rate of energy dissipation in the corona.  The ratio of these two, $P/\dot{\Phi}$, is a current comparable to the amount of current expected to flow along the boundary separating the emerged flux from the pre-existing flux overlying it.  We can generalize this relation to the overall corona in quiet Sun or in active regions.  Doing so yields estimates for the contribution to corona heating from magnetic reconnection.  These estimated rates are comparable to the amount required to maintain the corona at its observed temperature.
\end{abstract}

\section{Introduction}

There is still little consensus on what mechanism can be credited with supplying heat to the Sun's corona.  Among the most frequently invoked  candidates are dissipation of waves and magnetic reconnection.  Both processes are known to occur, but their relative contributions to heating has yet to be definitively quantified observationally.

There is extremely strong evidence that magnetic reconnection is occurring throughout the corona at some rate.  The coronal field is connected to photospheric flux concentrations which are, in all the best observations, surrounded by photosphere unconnected to the coronal field, if not entirely unmagnetized.  These flux concentrations move about, apparently at random, under the influence of granular and super-granular flows.  If the coronal 
field lines remained anchored to the same pair of footpoints over days or weeks, the coronal magnetic field would appear extremely tangled and complex.  The coronal  
field outlined in EUV images shows little sign of such tangling --- in fact, it appears smooth enough to have been ``combed''.   While it is still possible that complex tangling occurs at length scales below our present resolution \cite{Berger2009}, footpoint motions occur over all length scales and presumably so should the tangling.  There is little evidence for it on the the largest scales, which appear increasingly well fit by potential fields over time \cite{Schrijver2005}.
This fact gives a clear indication that coronal field lines are constantly being reconnected: uprooted from one footpoint and reattached to another.

Based on this reasoning we propose that magnetic field lines in a given portion of the corona are undergoing topological change at some rate $\dot{\Phi}$.  Two questions raised by this proposition are, firstly, how is topological change related to heating?, and secondly, what fraction of overall coronal heating can be attributed to topological change?  We address these two questions in the following.

\section{Heating from magnetic reconnection}

\begin{figure}[htb]
\centerline{\includegraphics[width=2.65in]{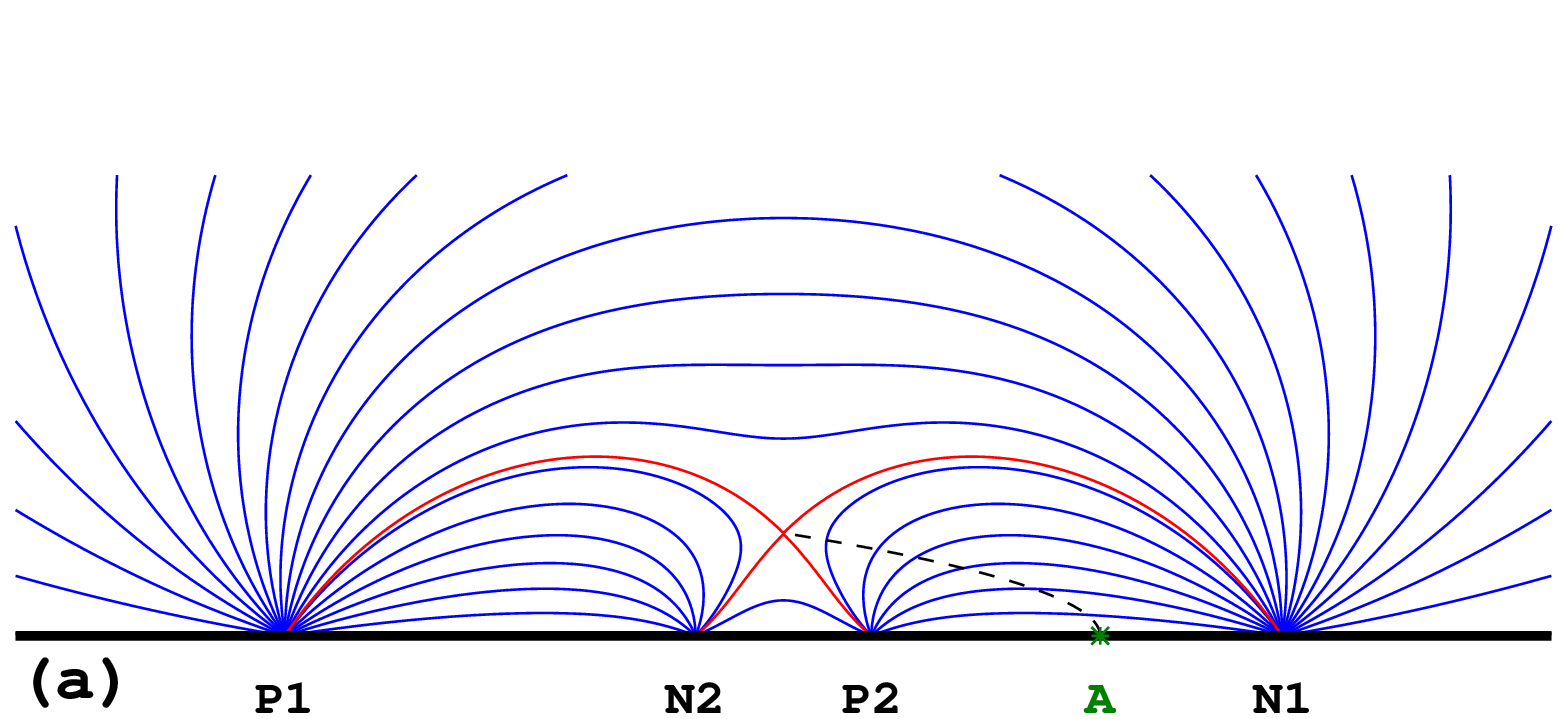}~~\includegraphics[width=2.65in]{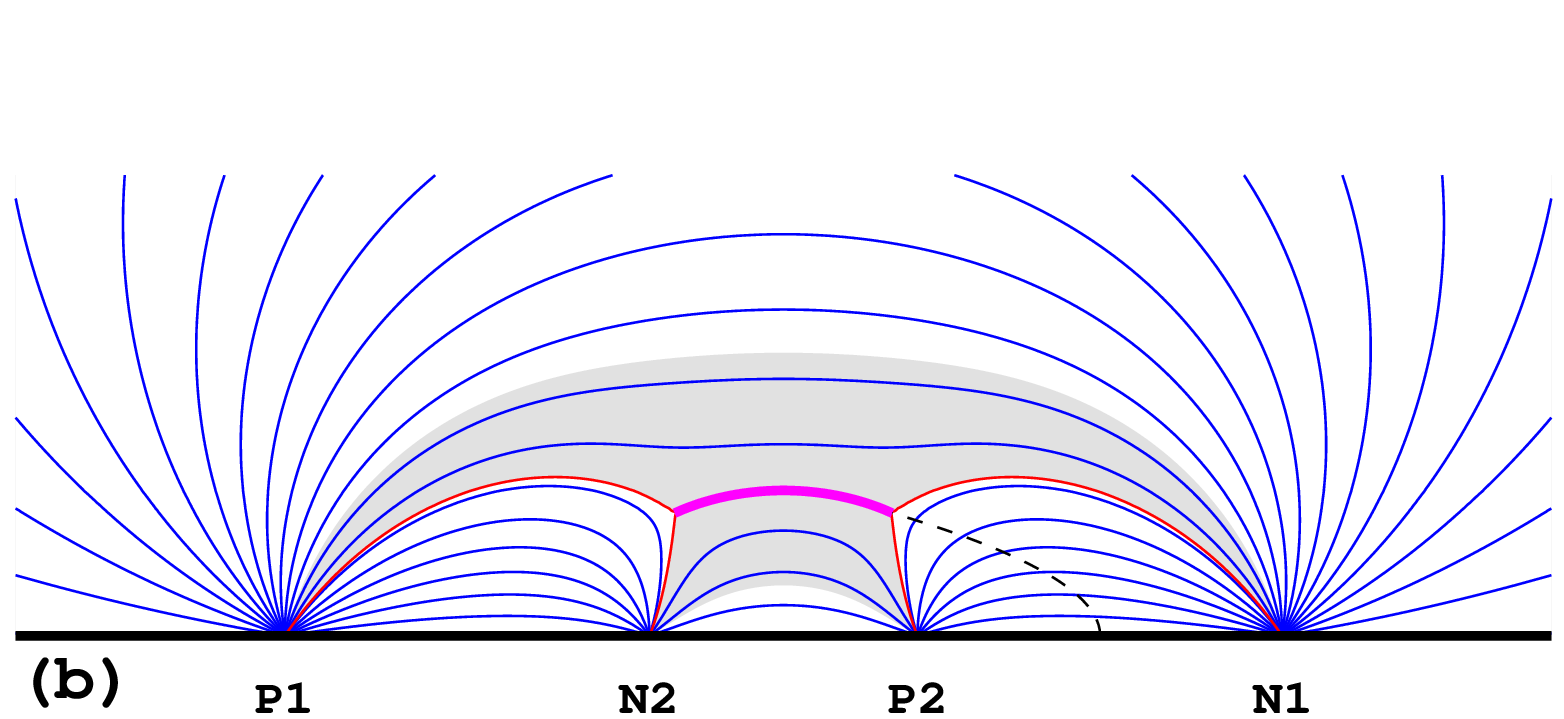}}
\centerline{\includegraphics[width=2.65in]{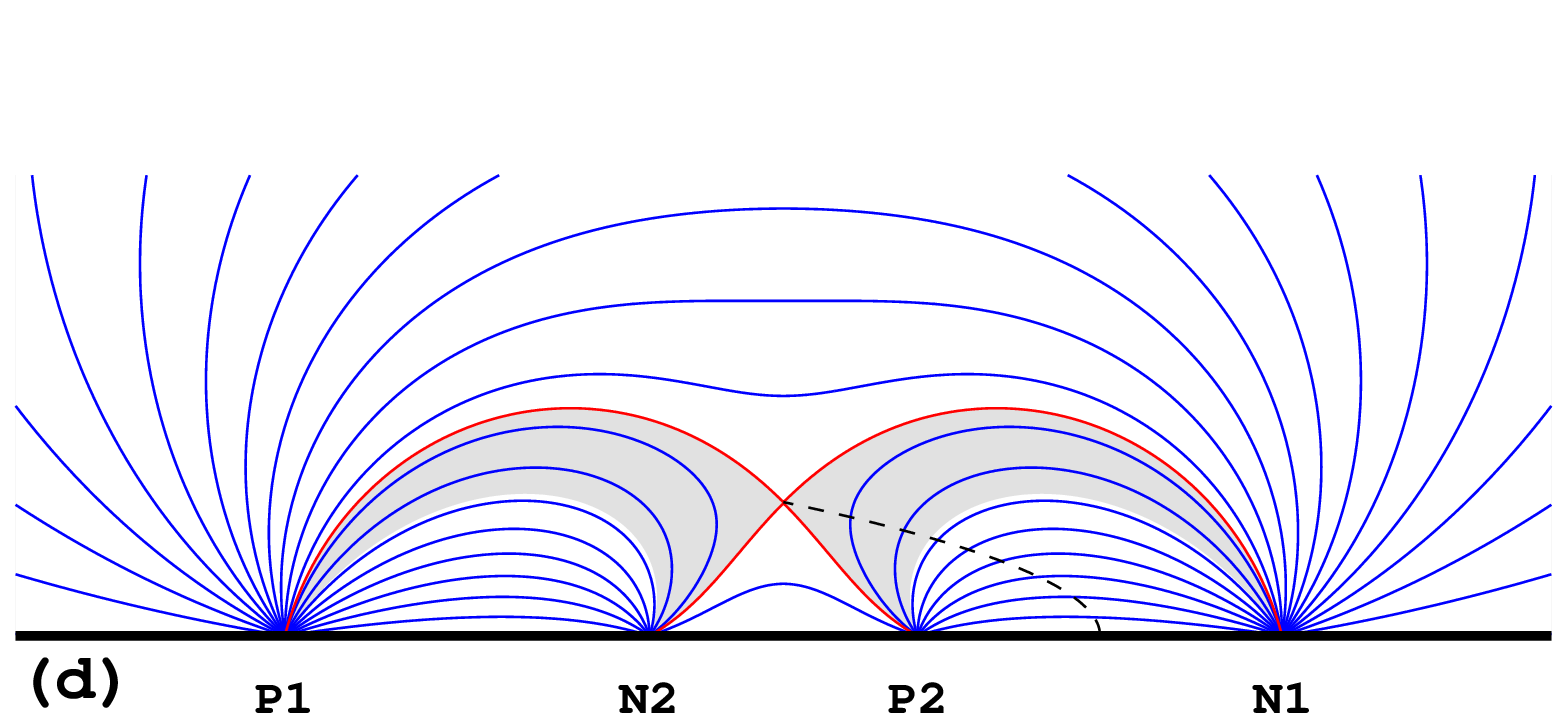}~~\includegraphics[width=2.65in]{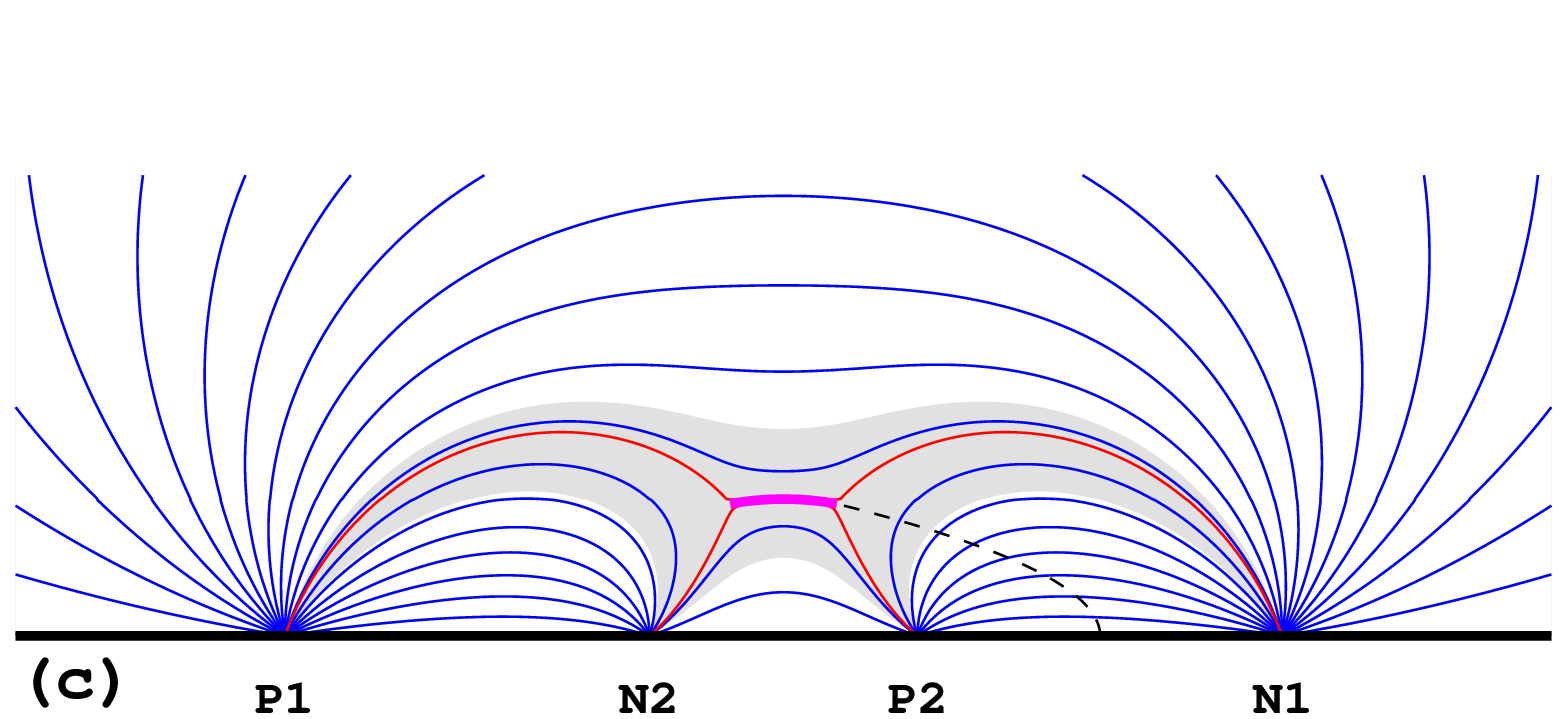}}
\caption{Stages, progressing clockwise from upper left, in reconnection following the emergence of the bipole beneath existing bipole.  Field lines are blue curves originating in sources P1, P2, N1, and N2, located in the photospheric boundary (black horizontal line).  (a) Shows the pre-emergence state, (b)--(d) show stages after emergence of bipole P2--N2 --- (b)--(d) have the same photospheric field.  (b) is the state before any reconnection,  ({c}) is after some reconnection and (d) is the state after complete reconnection.  Red lines show the separatrices, magenta curves are the current sheet.  The shaded region shows the flux which is reconnected to produce state (d) from state (b).  Dashed curves show a surface though which flux connecting P2 to N1 can be computed.}
	\label{fig:toon}
\end{figure}

To illustrate how the topological change of magnetic field lines could result in plasma heating we consider a simplified, two-dimensional model of flux emergence sketched in \fig\ \ref{fig:toon}.  In the initial state, \fig\  \ref{fig:toon}a, a small bipole, P2--N2, sits beneath a larger, existing bipole P1--N1, and the coronal magnetic field is potential: 
$\nabla\times{\bf B}=0$. The next three panels, \ref{fig:toon}b--\ref{fig:toon}d, show a state in which the small bipole has emerged further (P2 and N2 have separated and their fluxes have increased) while the outer bipole has not changed.  

In \fig\ \ref{fig:toon}d, where the coronal field is once again potential, the flux interconnecting old and new poles, P2 and N1, has increased.  The interconnecting flux is computed through a surface, depicted by a dashed curve, which extends from the coronal X-point to a point $A$ located on the boundary somewhere between P2 and N1.  Between \figs\ \ref{fig:toon}a and \ref{fig:toon}d this flux has increased by the addition of flux shown shaded in \ref{fig:toon}d.  According to Faraday's law this additional flux must have arisen through an electric field\footnote{We use cgs Gaussian units with the exception of electric field and current.  In the interest of brevity we use  for those, ${\bf E}=c{\bf E}_{\rm cgs}$ and $I=I_{\rm cgs}/c$, whose units are ${\rm G\cdot cm/s}$ and ${\rm G\cdot cm}$ respectively.}
\begin{equation}
  \dot{\Phi} ~=~ -\oint_{\cal C}{\bf E}\cdot d{\bf l} ~~,
  	\label{eq:Faraday}
\end{equation}
along the perimeter ${\cal C}$ of the surface.  If we assume no interconnecting field lines moved upward through the boundary at point $A$ (such motion occurred only within the region between P2 and N2, as flux emergence) then there is no electric field along the horizontal leg at $A$.  Instead, the entire electric field in integral (\ref{eq:Faraday}) occurs at the coronal X-point --- this is the reconnection electric field.

If the coronal field had remained entirely potential throughout the emergence the reconnection would have produced {\em no} coronal heating at all --- there would have been topological change in the absence of heating.  The rate of electromagnetic work on the plasma is
\begin{equation}
  P ~=~ I\, \dot{\Phi} ~~,
  	\label{eq:power}
\end{equation}
where $I$ is the current flowing along perimeter ${\cal C}$.  Without coronal current there is no electromagnetic work, and therefore no heating from the topological change.

Heating occurs only when the magnetic reconnection is slow enough to permit the accumulation of current at the X-point.  Figure \ref{fig:toon}b shows the case where no reconnection at all has occurred during the flux emergence.  The flux interconnecting P2 and N1 is therefore the same as before emergence and there is a current sheet separating the newly emerged from the pre-existing flux \cite{Heyvaerts1977}.  (The state with a current sheet is the one with the lowest magnetic energy subject to the single constraint on the interconnecting flux \cite{Longcope2001b}.)  Any electric field within the current sheet will increase the P2--N1 flux, taking \fig\ \ref{fig:toon}b to \fig\ \ref{fig:toon}c.  In this process the magnetic energy will decrease by doing work on the plasma, endowing it with either heat or bulk kinetic energy.  The rate of electromagnetic work is given by \eq\ (\ref{eq:power}) where $I$ is the current in the coronal current sheet at which the reconnection occurs.

\section{A case study: flux emergence in AR 11112}

The foregoing simplified example illustrates how flux emergence makes topological change clearly identifiable, and even quantifiable.   Longcope {\em et al.} \cite{Longcope2005} measured such a change for one active region (AR) emerging in the vicinity of an existing AR.  This allowed them to compute the rate of topological change 
$\dot{\Phi}$.   Unfortunately, the limited EUV and X-ray data available made it difficult to compute a heating power $P$ for this case (TRACE was observing at high cadence, but only in 171 \AA, and {\em Yohkoh} SXT had multi-filter partial frame images only during two intervals).  The observation was thus not ideally suited to understanding the relation between reconnection and heating captured in \eq\ (\ref{eq:power}).

\begin{figure}[htb]
\centerline{\includegraphics[width=5.3in]{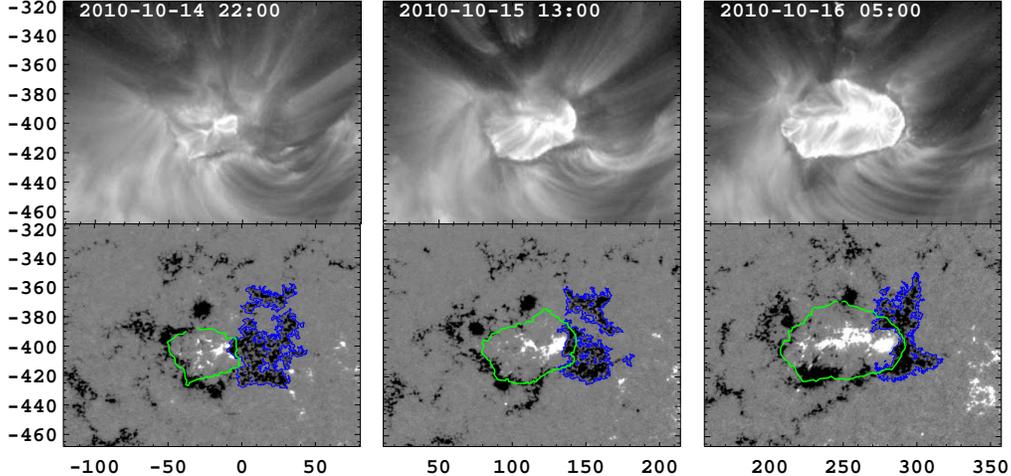}}
\caption{The emergence of a bipole into the negative polarity of AR 11112.  The top row shows AIA 211\AA\ images from three times during the emergence, plotted with a logarithmic grey-scale.  The bottom row shows HMI maps of $B_z(x,y)$ from the same times (on a grey-scale capped at $\pm500$ G).  The green curves outline the dome from the 211 \AA\ images, and the blue curves outline the old negative polarity, N1, to which reconnection occurs.}
	\label{fig:211}
\end{figure}

Recently  Tarr {\em et al.} \cite{Tarr2014} used the AIA and HMI instruments on SDO \cite{AIA,Scherrer2012} 
to observe a magnetic bipole emerging {\em within} an existing AR.  Figure \ref{fig:211} shows 211\AA\ images\footnote{The 211\AA\ filter of AIA is primarily sensitive to Fe {\sc xiv}, whose peak formation temperature is 
$T\simeq 2\times10^{6}$ K, in ionization equilibrium \cite{AIA}.} 
from AIA of the emerging bipole (top row) along with radial field maps from HMI (bottom row).  The EUV images clearly show a dome of flux anchored in the newly-emerged positive polarity, which we call P2 by analogy with \fig\ \ref{fig:toon}.  The newly emerged negative polarity has moved to the southeast, but the dome clearly includes negative polarity to the west: old flux, outlined in blue in the bottom row of \fig\ \ref{fig:211}, and hereafter called N1.  The distinction between old flux (N1) and new flux (N2) is made by tracking the evolution of the magnetograms, first automatically and then adjusting manually \cite{Tarr2012,Tarr2014}.  This introduces the largest source of error into our calculation.  One indication of its 
magnitude is that the total signed flux attributed to newly emerged flux remains constant within 10\% of 
$\Psi_{2}$ over the primary emergence period --- Oct.\ 15.

\begin{figure}[htb]
\centerline{\includegraphics[width=5.5in]{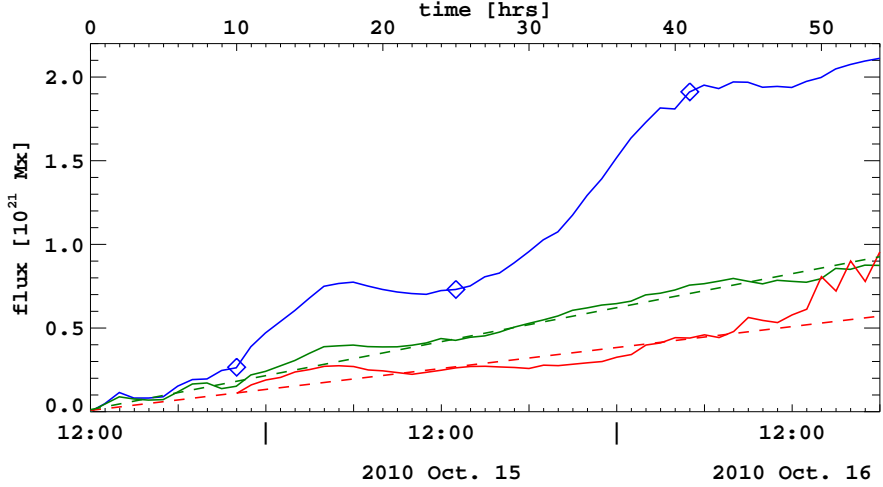}}
\caption{The fluxes from the emerging bipole over time.  The blue curve shows the total flux in the positive polarity P2 in units of $10^{21}$ Mx.  Diamonds mark the times of the panels in \fig\ \ref{fig:211}.  The red curve is $\Phi_r(t)$ found from integrating the portion of the old negative polarity within the dome.  The green curve is the amount of reconnected flux in a potential field, $\Phi_r^{\rm (v)}(t)$.  Dashed lines are linear fits, both intersecting zero at $11:15$ on Oct.\ 14.}
	\label{fig:phi_ar}
\end{figure}

The perimeter of the 211\AA\ dome was traced (manually) at one hour intervals and mapped onto the HMI radial field maps (green curves).  The total unsigned flux within this boundary remains constant to with 10\% of $\Psi_{2}$, 
suggesting that the boundary accurately identifies a closed magnetic system \cite{Tarr2014}.
The region of old negative flux (N1) lying within the dome perimeter was taken to be reconnected flux, analogous to the P2--N1 flux in \fig\ \ref{fig:toon}.  The integral of $B_z$ over the overlapping region gives the reconnected $\Phi_r(t)$ shown as a red curve in in \fig\ \ref{fig:phi_ar}.  A crude linear fit to this (dashed line) rises at $\dot{\Phi}_r\simeq2.9\times10^{15}$ Mx/s (Tarr {\em et al.} \cite{Tarr2014} fit with a line starting later and thus found a slightly larger slope).  

This value can be compared to that from a potential field extrapolation from the same radial field.  A magnetic charge topology (MCT) extrapolation replaces each region by a point source in order to tally the total flux connecting every pair of regions \cite{Longcope1996d}.  The flux connecting the emerged positive (P2) to the existing negative (N1), outlined by blue contours in \fig\ \ref{fig:211}, called $\Phi_r^{\rm (v)}(t)$, is shown as a green curve in \fig\ \ref{fig:phi_ar}, and rises at 
$\dot{\Phi}_r^{\rm (v)} \simeq 4.7\times10^{15}$ Mx/s (dashed line). This can be taken as an upper bound on the reconnected flux, in the same way that the potential field in \fig\ \ref{fig:toon}d included more inter-connecting flux than the states with current, \fig\ \ref{fig:toon}b or  \ref{fig:toon}c.

\begin{figure}[htb]
\centerline{\includegraphics[width=5.5in]{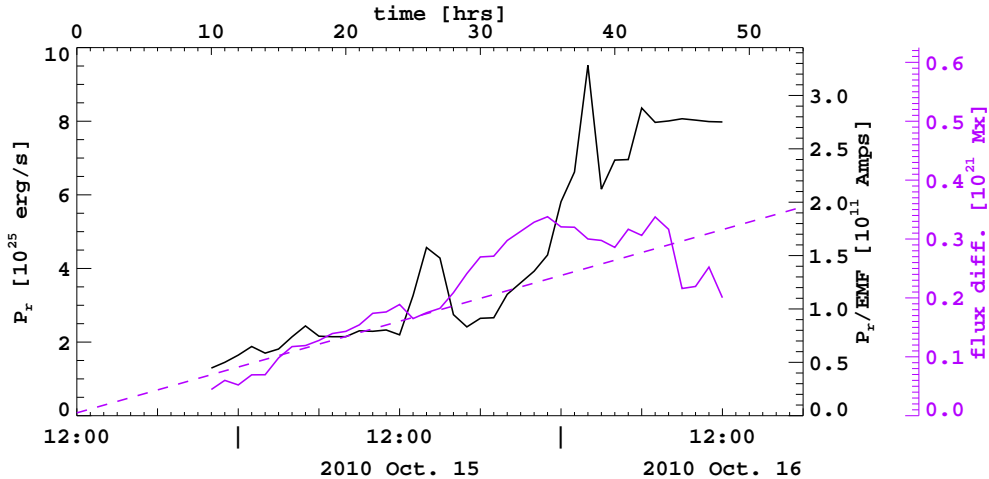}}
\caption{The radiated power $P_r$ from the dome plasma computed from {\em Hinode} XRT filter ratios.  The power is a black curve read in units of $10^{25}$ erg/s against the left axis.  The same curve is scaled by $\dot{\Phi}_r\simeq2.9\times10^{15}$ to yield a current, read from the inner right axis in units of $10^{11}$ Amps.  The flux difference, defined by \eq\ (\ref{eq:dPhi}), is plotted in violet after being scaled by the factor $\xi=1.6\times10^5$ G cm/s.  It can be read from the outer right axis in units of $10^{21}$ Mx.}
	\label{fig:pwr_ar}
\end{figure}

The emerging region was also observed by the X-ray Telescope (XRT) on {\em Hinode} \cite{Hinode}.  A ratio of images from Ti-poly and Al-mesh filters is used to derive the temperature and emission measure of the plasma within the dome \cite{Narukage2011,Tarr2014}.  This is then used with the total radiative loss function from Klimchuk and Cargill \cite{Klimchuk2001b} to derive the power, $P_r$, radiated by the dome plasma, shown as a black curve in \fig\ \ref{fig:pwr_ar}.  To generate this power entirely by the reconnection at the observed rate $\dot{\Phi}_r\simeq2.9\times10^{15}$, would require current $I=P_r/\dot{\Phi}_r$, in accordance with \eq\ (\ref{eq:power}).  This value, found by reading the black curve from the right axis, rises to over 
$I\simeq 3\times10^{11}$ Amps over the 50 hours of emergence.

In a simplistic view of this process, the emerging flux reconnects to its surroundings in order to reduce its magnetic energy, and thereby approach a potential field.  The reconnection is therefore driven by the difference in fluxes between the actual interconnecting flux and that in a potential field
\begin{equation}
  \Delta\Phi ~=~ \Phi_r^{\rm (v)} ~-~ \Phi_r  ~~,
  	\label{eq:dPhi}
\end{equation}
which is the separation between between the green and red curves in \fig\ \ref{fig:phi_ar}.  This flux difference is plotted in violet in \fig\ \ref{fig:pwr_ar}.  In order to plot it with $P_r$, the flux difference $\Delta\Phi$ is multiplied by the scaling factor $\xi=1.6\times10^5$ G cm/s, chosen to bring the two curves into approximate alignment in the plot.  (A possible 
interpretation of this empirical scaling factor is suggested below.)

If the reconnection were occurring across one or more current sheets created only in response to the flux discrepancy 
$\Delta\Phi$ then we would expect a relation between $\Delta\Phi$ and $I$.  The simplest relation would be a linear one
\begin{equation}
  I ~= {\Delta\Phi\over{\cal L}} ~~,
  	\label{eq:induct}
\end{equation}
where ${\cal L}$ is formally a self-inductance, although the analysis so far has no current path with which to associate it 
(one possibility is suggested below).  Equating this current with the value, 
$P_r/\dot{\Phi}_r$, derived from the power, yields an expression for the inductance
\begin{equation}
  {\cal L} ~=~ {\dot{\Phi}_r\,\Delta\Phi\over P_r} ~\simeq~ {\dot{\Phi}_r\over\xi} ~=~ 1.8\times 10^{10}\, {\rm cm} ~~,
  	\label{eq:L_ar}
\end{equation}
after using the values quoted above.

A linear relationship like \eq\ (\ref{eq:induct}) obtains in the simple two-dimensional model of \fig\ \ref{fig:toon}: the current in the single, global sheet is proportional to the amount of grey, unreconnected flux underneath it \cite{Longcope2004}. 
In a three-dimensional model with isolated sources, the topological role of the X-point is played by a separator, and a sheet will form there carrying a current proportional the the flux difference $\Delta\Phi$ \cite{Longcope1996}.  In that model a separator loop of length $L$ has self-inductance is ${\cal L}=4\pi\, L$ times a logarithmic factor typically of order unity.  

\begin{figure}[htb]
\centerline{\includegraphics[width=2.2in]{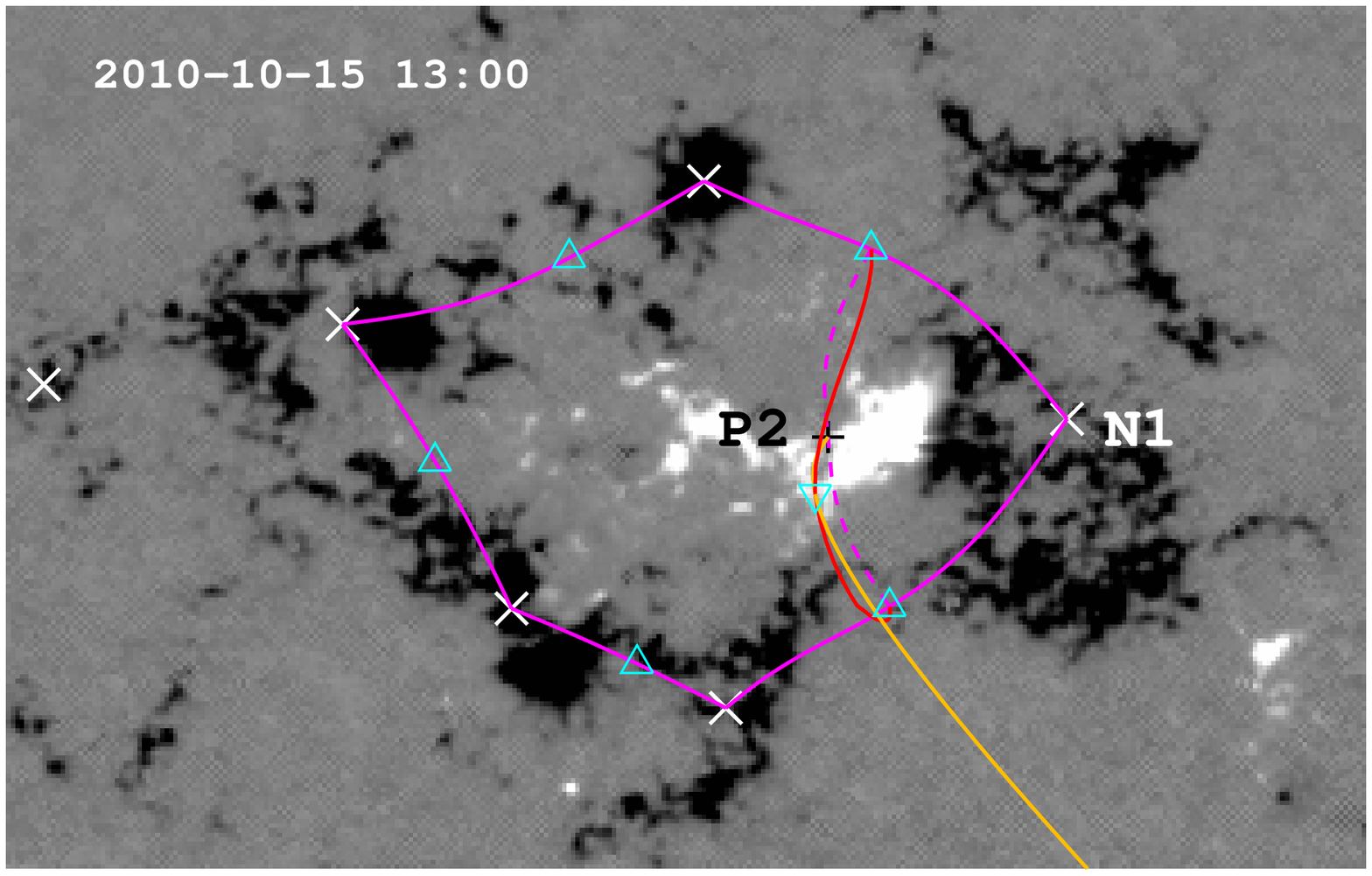}~~~\includegraphics[width=2.9in]{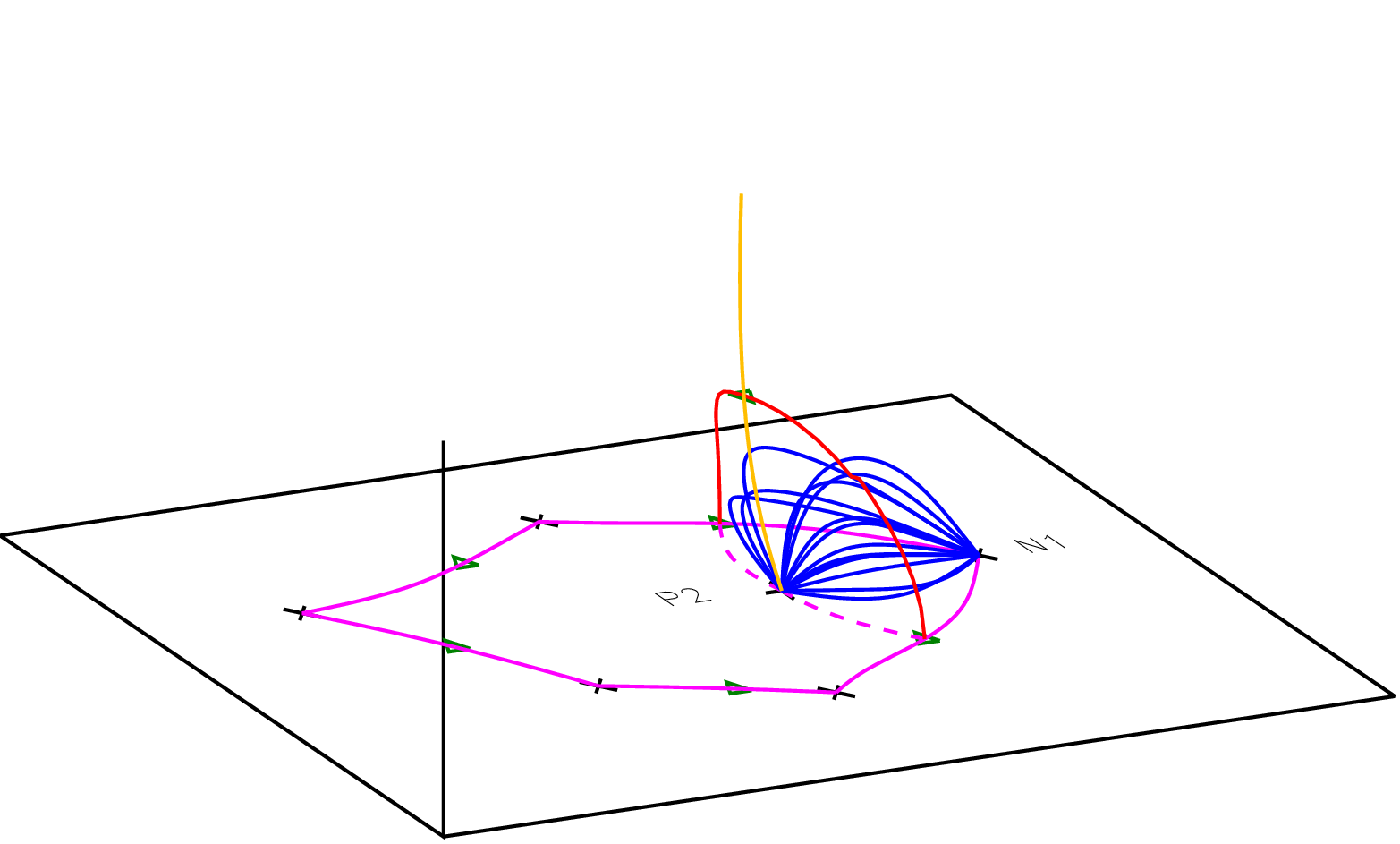}}
\caption{The MCT model of the field at 2010 Oct. 15 13:00 (the same field as the middle panel of \fig\ \ref{fig:211}).  The left shows the points source white $\times$s and black $+$s on top of the HMI radial field map (grey scale).  Photospheric negative nulls are indicated with cyan $\triangle$ and the coronal positive null with a cyan $\nabla$.  Magenta solid curves show the spine lines from the photospheric nulls.  These form a closed curve in the photosphere which is base of the separatrix dome --- the fan surface of the coronal null.  The solid orange curves show the two spines from the coronal null (one terminates at P2 the other connects to a positive source far away).  The dashed magenta curves are the bottoms of the separatrix surfaces which intersect the dome surface along the the two separators (solid red) enclosing the P2--N1 field lines.  The right panel shows the same features in a projection, with vertical scale exaggerated for clarity.  Nulls are green rather than cyan and a selection of P2--N1 interconnecting field lines are shown in blue.}
	\label{fig:mct}
\end{figure}

The MCT model of the emerging bipole, shown in \fig\ \ref{fig:mct}, includes a positive coronal null point (triangle) 15 Mm above the positive polarity P2, whose fan surface forms a separatrix dome.  The base of this surface is a ring of spines (solid magenta curves) linking the negatives sources to negative photospheric nulls, shown as green and cyan $\triangle$s.  These nulls are linked to the coronal null by separators lying within the separatrix dome.  Two of these separators, shown in red in \fig\ \ref{fig:mct}, form a loop enclosing the reconnected flux linking P2--N1 (blue field lines).  The current sheet will form along this loop, in analogy to that in \fig\ \ref{fig:toon}.  The separators are 24 Mm and 27 Mm long, creating a loop of total length $L=51$ Mm, comparable to the distance inferred from the self-inductance in \eq\ (\ref{eq:L_ar}), although it would seem the logarithmic factor in this case is $\simeq 3.6$.

The rate of observed magnetic reconnection, $\dot{\Phi}=2.9\times10^{15}$ Mx/sec, is equivalent to an electromotive force (EMF) of 29 Megavolts.  If this EMF were from a simple loop, then the mean electric field, $\dot{\Phi}/L\simeq 0.6$ V/m, would be about one hundred times greater than the Dreicer field.  We do not believe that such a large electric field could actually be present.  Aside from theoretical difficulties implied by such an electric field, there is absolutely no evidence of particles having energies even close to 29 MeV.  This suggest to us that the electric field within the reconnecting current sheet is far more complex than a simple closed loop.  The EMF might, instead, be built from numerous parallel reconnection events transferring reconnected flux through the separator loop.

This exercise provides one example where the reconnection rate, $\dot{\Phi}$, can be measured along with the heating power, or at least the portion $P_r$ lost through radiation.  The values found from these observations are in conformance with theoretical models in which reconnection produces the heating.  In particular, the ratio $P_r/\dot{\Phi}$, corresponds to a current which can be understood as that resulting from the emergence and accumulation of magnetic free energy.

Having related reconnection to heating in a single case, we take up the task, in the next section, of applying the relation to a more general setting: heating the entire corona.  Topological change can be quantified observationally in cases of small flux elements interacting in the quiet Sun.  The heating from these interactions can be used to predict a mean coronal heating rate from quiet Sun reconnection.  We apply a similar technique to predict heating from reconnection in a active region.

\section{A statistical view of reconnection heating}

To estimate the net contribution of reconnection to coronal heating we consider a collection of photospheric flux concentrations, of mean flux $\Phibar$, randomly distributed with areal density 
$\bar{n}$.  To remain in steady state one concentration must completely reconnect in the time $\tau_{\rm ex}$ that it takes it move into the position of a neighboring concentration.  If it moves randomly with typical photospheric velocity $v_{\rm ph}$, this  exchange time will be
\begin{equation}
  \tau_{\rm ex} ~\sim~ {1\over\, \sqrt{\bar{n}}\, v_{\rm ph}} ~=~ {\sqrt{\Phibar\,/\bar{B}}\over v_{\rm ph}} ~~,
  	\label{eq:tau_rc}
\end{equation}
where $\bar{B}=\bar{n}\,\Phibar$ is the mean flux density of one sign of concentrations.  (At this point we treat unipolar and mixed-polarity distributions simultaneously.)  The rate of flux
transfer into or out of domains anchored to a single concentration is therefore
\begin{equation}
  \dot{\Phi} ~=~ {\Phibar\over\tau_{\rm ex}} ~\sim~ v_{\rm ph}\, \sqrt{\bar{B}\,\Phibar} ~~.
\end{equation}
According to \eq\ (\ref{eq:power}), flux transfer across a current sheet with typical current $\bar{I}$ will do electromagnetic work
\begin{equation}
  P~=~ \bar{I}\dot{\Phi} ~\sim~ v_{\rm ph}\, \bar{I}\, \sqrt{\bar{B}\,\Phibar} ~~,
  	\label{eq:dPhi_pow}
\end{equation}
on the plasma occupying the coronal field anchored to that flux element.
The heat flux deposited into the entire corona is therefore
\begin{equation}
  F_H ~\sim~ \bar{n}\, P ~=~ v_{\rm ph}\, \bar{I}\, \bar{B}^{3/2}\,\Phibar\,^{-1/2} ~~,
  	\label{eq:FH_stat}
\end{equation}
per photospheric surface area.  When the coronal field evolves quasi-statically the current, $\bar{I}$, will be independent of velocity so the heat flux scales linearly with velocity.  This is natural for a quasi-static heating process which must release a fixed amount of energy from a given displacement, independent of how rapidly the displacement is made.

\subsection{Quiet Sun}
The quiet Sun corona lacks large-scale structure, and therefore contains current sheets only between neighboring flux concentrations.  These sheets extend a distance comparable to the separation between elements, $L\sim \bar{n}^{-1/2}$.  They are not driven by large-scale forces, but rather exist to compensate for an unreconnected 
flux which is some fraction $f$ of the entire element.  Complete flux exchange is therefore assumed to occur in $1/f$ distinct reconnection events.  (The value of $f$ will depend on the details of the reconnection process, and is considered a free parameter at this time.)  The current will therefore scale as
\begin{equation}
  I ~\sim~ {f\, \Phibar\over L} ~\sim~ f\,\sqrt{\bar{B}\,\Phibar} ~~.
  	\label{eq:Ibinary}
\end{equation}
Substituting this into expression (\ref{eq:dPhi_pow}) gives the heating power into a single element
\begin{equation}
  P~\sim~  f\, v_{\rm ph}\, \bar{B}\,\Phibar ~~.
  	\label{eq:Pxbp}
\end{equation}
It is evident that little heating power results when reconnection occurs relatively smoothly, in very small 
steps (i.e.\ $f\to0$). 

In a survey of quiet Sun magnetograms from MDI Longcope and Parnell \cite{Longcope2009} found the isotropized Fourier power spectrum and the kurtosis of the magnetograms could be matched by discrete flux elements with mean size 
$\Phibar=1.0\times10^{19}$ Mx distributed with an areal density of $\bar{n}=7.0\times 10^{-19}\,{\rm cm}^{-2}$.   Taking their random velocity to be $v_{\rm ph}=250$ m/s \cite{Hagenaar2001}
means they would exchange their flux in $\tau_{\rm ex}=13$ hours.  Close {\em et al.}  \cite{Close2005} used a potential field extrapolation from a sequence of from MDI quiet Sun magnetograms to compute the reconnection between flux elements.  They found a reconnection recycling time only one tenth the value derived above.  This faster reconnection might be due to a faster motion, or from reconnection sequences more complex than the binary case considered here.  In any event, we take this as evidence that ours is a conservative estimate of the quiet Sun reconnection rate.  According to the values from Longcope and Parnell, the mean flux density is $\bar{B}=\bar{n}\,\Phibar=7.0$ G in typical quiet Sun.

\begin{figure}[htb]
\centerline{\includegraphics[width=4.0in]{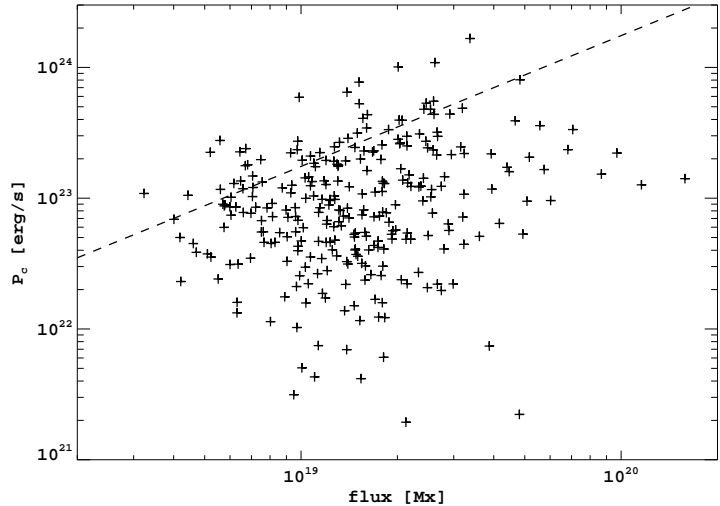}}
\caption{Power (conductive) {\em vs.} flux for a survey of 285 XPSs from Longcope {\em et al.} \cite{Longcope2001}.  Conducted power, $P_{\rm cond}$, from \eq\ (\ref{eq:Pcond}), is plotted against the flux from an average of both poles in the bipole.  The dashed line shows
\eq\ (\ref{eq:Pxbp}).}
	\label{fig:xbps}
\end{figure}

Longcope {\em et al.} \cite{Longcope2001} surveyed X-ray bright points (XBPs) using MDI magnetograms and EIT images in two wavelengths.  They used the EIT data to derive the emission measure and temperature of the XBP from which they computed the total radiated power.  The same data can, however, be used to compute the power conducted downward from the XBP
\begin{equation}
  P_{\rm cond} ~=~{\kappa\, T\over (L_{\rm bp}/2)^2} \, V ~~,
  	\label{eq:Pcond}
\end{equation}
where $\kappa = \kappa_0T^{5/2}$ is the Spitzer thermal conductivity ($\kappa_0=10^{-6}$ in cgs units).
It is probable that heat conduction occurs by classical means in events of such small energy flux. 
The separation between the magnetic concentrations is $L_{\rm bp}$ and  the short and long axes of the EIT feature, 
$\ell_1$ and $\ell_2$, are used to estimate the coronal volume at $V=(4\pi/3)\ell_1^2\ell_2$.  Figure \ref{fig:xbps} shows the conducted power from the 285 XBPs in the survey {\em vs.} the mean flux of their photospheric concentrations.  The dashed line shows \eq\ (\ref{eq:Pxbp}) with $f=0.1$, $v_{\rm ph}=250$ m/s and $\bar{B}=7$ G, as described above.  The line lies along the top of the observed distribution, suggesting that our values of velocity, mean flux density, or $f$, are slightly larger than reality: their product may be excessive by about a factor of three.

The average reconnection heat flux to the quiet Sun is found using the current for binary interactions, \eq\ (\ref{eq:Ibinary}), in expression (\ref{eq:FH_stat}).  The result
\begin{equation}
  F_H^{\rm (QS)} ~\sim~ f\, v_{\rm ph}\, \bar{B}^2 ~~,
  	\label{eq:FH_QS}
\end{equation}
is, remarkably, independent of element size.  The values quoted above yield a heat flux $F_H=10^5\,{\rm erg\,cm^{-2}\,s^{-1}}$, consistent with the heat flux inferred for quiet Sun regions \cite{Aschwanden2001b}.

\subsection{Active region heating}

Unlike the quiet Sun, an active region has global currents driven on large scales.  A force-free magnetic field will have a global current density $J =\alpha\bar{B}$. The field is anchored to discrete elements but we will assume that current does not flow into any of the photospheric elements.  Instead the coronal current will be carried along separatrices where it can close along the chromospheric canopy between concentrations \cite{Longcope2002}.   The result will be coronal current sheets confined to the peripheries of cells.  The peripheral sheet associated with a single element will carry all the current which would have otherwise entered that cell
\begin{equation}
  I ~\sim~ J\, {\Phibar\over\bar{B}} ~=~ \alpha\, \Phibar ~~.
\end{equation}
Flux transfer into or out of the domain anchored to this concentration must occur across this current sheet, thereby giving rise to electromagnetic work and heating.   Substituting this into \eq\ (\ref{eq:FH_stat}) gives an active region heating rate
\begin{equation}
  F_H^{\rm (AR)} ~\sim~ \alpha\, v_{\rm ph}\, \bar{B}^{3/2}\,\Phibar\,^{1/2} ~~.
  	\label{eq:FH_AR}
\end{equation}
The differences between this expression and the quiet Sun heating rate, \eq\ (\ref{eq:FH_QS}), are due to the different natures of the currents across which the reconnection occurs: generated globally {\em vs.} locally.  One implicit consequence of this difference is that since the AR current is not driven by a flux discrepancy, 
reconnection will not necessarily reduce it.

The active region heating is also greater due to the greater density of flux concentrations: typically $\bar{B}\sim100$ G in plage.  Assuming the same mean concentration size, $\Phibar=10^{19}$ Mx, with the same random velocity, 
$v_{\rm ph}=250$ m/s, and taking a typical twist $\alpha\sim 10^{-10}\,{\rm cm^{-1}}$ \cite{Pevtsov1995,Leka1999b}, yields a heat flux $F_H^{\rm (AR)}\sim 10^7\,{\rm erg\,cm^{-2}\,s^{-1}}$, 
comparable to that expected of active regions \cite{Withbroe1977,Aschwanden2001b}.

\section{Discussion}

If heating power $P$ were generated by magnetic field lines undergoing topological changes, i.e.\ reconnection, at a rate $\dot{\Phi}$, then the ratio of these rates $P/\dot{\Phi}$ would have units of electric current.  The foregoing showed examples where that relation arose from an actual current in the coronal magnetic field.  In those examples the reconnection electric field does electromagnetic work on the plasma at a rate $P=I\,\dot{\Phi}$.  This is how magnetic reconnection might heat the solar corona.

From that basic scenario we derived scaling laws quantifying the heat that could arise as magnetic elements move randomly over the photosphere, and coronal reconnection occurs to keep the field there from becoming excessively tangled.  The process described is, at its root, the same one used by Parker to arrive at the reconnection heat flux 
$F_H\sim v_{\rm ph}B^2_z\tan\theta$, for field lines pushed to an angle $\theta$ from their 
potential state \cite{Parker1983b}.  Indeed, the relation found here for quiet Sun heating, \eq\ (\ref{eq:FH_QS}), 
has the same scaling.  In our expression, however, $\bar{B}$ is the density of small-scale flux elements, rather than a vertical magnetic field strength.  In point of fact, we tacitly assume the concentrations have local field strength 
$B_z\gg\bar{B}$ in order that they be distinct entities.  Notably, the local field strength does not enter our estimate of heating, since heating is assumed to occur in the corona rather than within the photospheric concentration.

Expression (\ref{eq:FH_QS}), and its interpretation,  matches the one derived by Longcope and Kankelborg\cite{Longcope1999b} under the consideration of interacting mixed-polarity magnetic elements.  That earlier derivation used an interaction model in which sources disconnected from an overlying background field in order to reconnect with one another \cite{Longcope1998b}.  The derivation proposed here assumes only that sources exchange flux across current sheets created through their binary interaction.

We recognize that coronal currents have different origins in ARs than in the quiet Sun: the former being globally rather than locally driven.  This recognition led us to  a reconnection heat flux, \eq\ (\ref{eq:FH_AR}), scaling as $F_H\sim\bar{B}^{3/2}$ in ARs.  This expression followed, once again, from reconnection between small magnetic elements, of a single polarity, composing the AR.  Priest {\em et al.} \cite{Priest2002} treated a similar scenario in their {\em flux-tube tectonics} model, but found a heat flux scaling similar to Parker's (their equation [53]).  They assumed that currents between interacting elements were driven locally, and therefore obtained the same scaling as we did in our quiet Sun derivation, which made the same assumption.

The weaker dependence we find on photospheric flux concentration, $F_H\sim \bar{B}^{3/2}$, may be in better agreement with observations.  Some studies have synthesized EUV and X-ray images from coronal equilibrium fields with {\em ad hoc} heating fluxes $F_H\sim B^{\nu}$ \cite{Schrijver2004,Lundquist2008b}.  These found the most reasonable matches 
with observations when $\nu$ was near unity.  

Heat flux, \eq\ (\ref{eq:FH_AR}), also scales with the large-scale twist, 
$\alpha$, in the AR, since that is assumed to be the source of current.  This is a a natural dependence for any model since more twisted fields contain more magnetic energy to be released by reconnection.  A statistical study of active regions by Fisher {\em et al.} \cite{Fisher1998} found no evidence for a scaling with the mean $\alpha$ of an AR.  This could mean that reconnection is not the primary source of heat for ARs, or that the currents across which reconnection occurs are driven on scales smaller than the entire AR.

Whether it supplies the majority of coronal heat or not, magnetic reconnection is clearly generating heat in the corona at some rate.  We have quantified this in \eq\ (\ref{eq:power}) as the rate of electromagnetic work attributable to topological change.  This is not necessarily Joule heating.  In several more detailed reconnection models this work is transferred primarily into bulk kinetic energy either in supersonic reconnection outflows, which generate heat through shocks
\cite{Petschek1964,Soward1982,Longcope2009} or in MHD waves which ultimately damp to generate heat 
\cite{Longcope2007e,Longcope2012}.  In view of the latter option, reconnection is not necessarily an alternative to heating by waves; rather it is a potential source of waves to heat the corona.

\bigskip

This work was partially supported by a grant from the NSF/DOE plasma sciences program.
We gratefully acknowledge two anonymous referees whose comments helped us improve the manuscript.


\end{document}